\documentclass[preprint,onecolumn,amsmath,amssymb]{revtex4}
\usepackage{CJK}
\usepackage{graphicx}
\usepackage{dcolumn}
\usepackage{bm}
\usepackage{times}
\usepackage{amsmath}
\usepackage{graphicx}
\usepackage{subfig}
\usepackage{caption}
\usepackage{epstopdf}
\usepackage{color}
\usepackage[abs]{overpic}

\begin{document}
\title{ Geometric phase and topological phase diagram of the one-dimensional $XXZ$ Heisenberg spin chain in a longitudinal field}
\author{Yi Liao$^{1,2}$, Xiao-Bo Gong$^{3}$, Chu Guo$^{4}$ and Ping-Xing Chen$^{1,2,}$}
\email{pxchen@nudt.edu.cn}
\affiliation{$^1$ Department of Physics, National University of Defense Technology, Changsha, 410073, China\\
$^2$ Interdisciplinary Center for Quantum Information, National University of Defense Technology, Changsha, 410073, China\\
$^3$ Yunnan Observatory and Key Laboratory for the Structure and Evolution of Celestial Objects, Chinese Academy of Sciences, Kunming, 650011, China\\
$^4$ Henan Key Laboratory of Quantum Information and Cryptography, SSF IEU, Zhengzhou, 450001, China}
\begin{abstract}
  In this paper, we determine the geometric phase for the one-dimensional $XXZ$ Heisenberg chain with spin-$1/2$, the exchange couple $J$ and the spin anisotropy parameter $\Delta$ in a longitudinal field(LF) with the reduced field strength $h$. Using the Jordan-Wigner transformation and the mean-field theory based on the Wick's theorem, a semi-analytical theory has been developed in terms of order parameters which satisfy the self-consistent equations. The values of the order parameters are numerically computed using the matrix-product-state(MPS) method. The validity of the mean-filed theory could be checked through the comparison between the self-consistent solutions and the numerical results. Finally, we draw the the topological phase diagrams in the case $J<0$ and the case $J>0$.
  \end{abstract}
\maketitle

\section{Introduction}
The phase transition plays an important role in quantum physics\cite{Yeomans}. It includes the Ginzburg-Landau(GL) phase transition and topological phase transition. { In the Ehrenfest's classification}, based on the continuity of the order parameter on the phase boundary, the GL phase transition has the first, second and $n$th-order transition. The fluctuation described in the GL phase transition theory could be thermodynamic or quantum. The GL phase transition thus falls into two categories: the thermodynamic(or classical) and quantum phase transition\cite{Sachdev,Fradkin2013}. The another kind of phase transition is called the topological phase transition where there exists the corresponding topological order parameter, such as the Berezinskii-Kosterlitz-Thouless (BKT)transition\cite{Berezinskii,Kosterlit}. { In the topological phase transition, the local fluctuation description does not work.}  It is formally classified into two types: the classical and quantum topological transition. Here, the terminology "quantum" only means that the topological transition only occurs near the zero temperature. By way of illustration, the well-known BKT phase transition is a kind of transition of infinite order and a classical topological transition. And the quantum BKT transition has been investigated\cite{Jensen}.

The geometric phase, or Berry phase(BP), a typical topological order parameter, entered the lexicon of physics about 30 years ago\cite{Berry,Liao,Liao2020}. Since then, numerous applications and experimental confirmations of this phase have been found in various physical systems \cite{Mead,Resta,Yarkony,Garg}. The BP can be exploited as a tool to detect topological phase transition. And if there exists nonzero BP in the system, it means that there exists parity and time inverse symmetry breaking for the ground state. The relationship between BP and topological transition in quantum system has been notoriously discussed in many literatures\cite{Sachdev,Bohm}.

The geometric phase and topological property in the spin systems, such as Ising model and $XY$ model, are important topics \cite{Bohm,Fradkin2013,Suzuki,Fradkin1978,Carollo,Zhu,Chen,Ma,Wang,Jalal,Cheng}. The $XXZ$ Heisenberg model is a more complex spin model used in the study of critical points and phase transitions of magnetic systems. In the condensed matter physics, the $XXZ$ Heisenberg model is also an important type of quantum dimer models to understand Bose-Einstein condensation. The quantum dimer magnet state is one in which quantum spins in a magnetic structure to form a  entangled singlet state. These entangled spins act as bosons and their excited states (triplons) can undergo Bose-Einstein condensation\cite{Giamarchi, Zapf}. The quantum dimer system was originally proposed as a mapping of the lattice Bose gas to the quantum anti-ferromagnet. There are some researches on quantum transition of the $XXZ$ Heisenberg model that is arguable one of the most fundamental examples of frustrated antiferromagnetism\cite{Yamamoto,Sellmann}. { Recently, R. Toskovic et al. have achieved the $XXZ$ model for researching the quantum criticality through manipulating single atoms\cite{Toskovic}. Their results present opportunities for further studies on quantum behaviours of many-body systems.} The phase diagram of the GL quantum phase transition which should be well-known drawn as a function of the reduced field strength $h$ and the spin anisotropy parameter $\Delta$ has been published in many systems. For example, on the triangular lattice,  Yamamoto et al. have drawn the quantum phase diagram using the cluster mean-field theory \cite{Yamamoto}. However, the phase diagram of the topological quantum transition in the $XXZ$ Heisenberg model have not been well drawn. Our paper aims at this for the one-dimensional $XXZ$ Heisenberg chain with spin-$1/2$ and exchange couple $J$ in a longitudinal field. Here the Berry phase is the corresponding topological order parameter.

The outline of this paper is as follows. In Sec.\ref{sect:Model}, with the method of the Jordan-Wigner transformation and the mean-field theory based on the Wick's theorem, we deal with the four-fermion interaction in the terms of some order parameters. In Sec.\ref{sect:MPS}, by means of the matrix-product-state(MPS) method£¬ we calculate these order parameters meaning the correlations between some creation and annihilation operators. The correlation plays a crucial role in determining the non-trivial BP. In Sec.\ref{sect:BP}, we map the Hamiltonian of one-dimensional $XXZ$ Heisenberg model in the momentum space to a two-level system where the BP has been well formulated. In terms of the order parameters, we obtain the formula of the BP.  Considering the validity of the mean-field theory, we draw the topological phase diagram in the paramagnetic systems where $J<0$  and the diamagnetic systems where $J>0$. In Sec.\ref{sect:Results}, we draw the conclusion and discuss our results related to the topological transition.

\section{One-dimensional $XXZ$ Heisenberg model in the longitudinal field}
\label{sect:Model}
The Hamiltonian of the one-dimensional $XXZ$ Heisenberg model in the longitudinal field reads
\begin{equation}
H_{_{XXZ}}\equiv J \tilde{H},\tilde{H}=h \overset{N}{\underset{i=1}{\sum}} S^z_i +\overset{N-1}{\underset{i=1}{\sum}}[\Delta(S^x_i S^x_{i+1}+S^y_i S^y_{i+1})+(S^z_i S^z_{i+1})].
\end{equation}
 Here, the notation $J$ is the exchange couple, the index $h$ is the reduced field strength and the index $\Delta$ is the spin anisotropy parameter. In the paper we only consider that lattice point number $N\rightarrow +\infty$, and the temperature $T\rightarrow 0$.  We introduce $S^{\pm}_i=S^x_i\pm jS^y_i,$ here $j$ denotes the imaginary unit satisfying $j^2=-1.$ Meanwhile we could make the Jordan-Wigner transformation which reads
\begin{eqnarray}
c_i&=&K_iS^-_i, c^+_i=S^+_iK^+_i,S^z_i=c_i^+c_i-\frac{1}{2};
\\ \nonumber
K_i&=&\exp[j\pi \overset{i-1}{\underset{m=1}{\sum}}S^+_mS^-_m], \{c_i,c_l^+\}=\delta_{il},\{c_i,c_l\}=\{c^+_i,c_l^+\}=0. \end{eqnarray}
Thus the reduced Hamiltonian reads
\begin{equation}
\tilde{H}=h \overset{N}{\underset{i=1}{\sum}} (c^+_ic_i-\frac{1}{2}) +\overset{N-1}{\underset{i=1}{\sum}}[\frac{\Delta}{2}(c^+_ic_{i+1}+c^+_{i+1}c_i)+(c^+_ic_i-\frac{1}{2})(c^+_{i+1}c_{i+1}-\frac{1}{2})].
\end{equation}
It also reads
\begin{equation}
\tilde{H}=\frac{(1-2h)N}{4}+ \overset{N}{\underset{i=1}{\sum}} (h-1)(c^+_ic_i) +\overset{N-1}{\underset{i=1}{\sum}}[\frac{\Delta}{2}(c^+_ic_{i+1}+c^+_{i+1}c_i)+(c^+_ic_ic^+_{i+1}c_{i+1})].
\end{equation}
Based on the Wick's theorem, ones could adopt a mean-field approximation which reads  \cite{Liao2020,Mahdavifar}
\begin{equation}
c_i^+c_ic_{i+1}^+c_{i+1}\approx -Z+R (c^+_{i+1}c_{i+1}+c^+_ic_i)+(C c^+_{i+1}c^+_i+C^*c_ic_{i+1})- (D c^+_ic_{i+1}+D^*c^+_{i+1}c_i).
\end{equation}
Here, the order parameters $ R \equiv<c^+_ic_{i}>=<S^z_i>+\frac{1}{2}; C\equiv<c_ic_{i+1}>;D\equiv <c^+_{i+1}c_i>.$
$Z\equiv R^2-|C|^2+|D|^2.$ The sign $|g>$ denotes the ground state of the system . And the sign $<O>\equiv <g|O|g>$ denotes the expectation of operator $O$ in the ground state. Further $D$ is a real number {as was shown} in our previous work \cite{Liao2020}. Therefore, the reduced Hamiltonian reads
\begin{eqnarray}
\tilde{H}&\approx& \frac{(1-2h-4Z)N}{4}+\underset{i=1}{\sum}[(h+2R-1)c_i^+c_i]
\\ \nonumber &+& \underset{i=1}{\sum}\{[(C c^+_{i+1}c^+_i+C^*c_ic_{i+1})]- [(D-\frac{\Delta}{2})c^+_ic_{i+1}+(D^*-\frac{\Delta}{2})c^+_{i+1}c_i)]\}
\\ \nonumber &\equiv& N A+ \underset{i=1}{\sum}B c_i^+c_i+ \underset{i=1}{\sum}[(Cc^+_{i+1}c_i^++C^*c_ic_{i+1})-(F c_i^+c_{i+1}+F^* c^+_{i+1}c_i)].
\end{eqnarray}
Here, $B\equiv 2R+h-1; F\equiv D-\frac{\Delta}{2}$ and $A\equiv\frac{1-2h-4Z}{4}$. And it is noticeable that the Wick's rule reads $|C|^2-|D|^2=<S^z_i S^z_{i+1}>-<S^z_i><S^z_{i+1}>.$  The sign $<S^z_iS^z_{i+1}>$ denotes the nearest-neighbor spin-spin correlation.  It will be proved in Sec.\ref{sect:BP} that $R$, $C$ and $D$ are independent of the position of the $i$-th lattice. { And the self-consistent equations about $R$, $C$ and $D$ will also be expressed in Sec.\ref{sect:BP}.}

One can switch to momentum space by the Fourier transformation when $N\rightarrow +\infty$, which reads
\begin{equation}
c_i=\frac{1}{\sqrt{2\pi}}\int c_k e^{jik}dk; c_k=\frac{1}{\sqrt{2\pi}}\int c_i e^{-jik}di.
\end{equation}
The reduced Hamiltonian $\tilde{H}$ reads
\begin{equation}
\tilde{H}=N A+\int_{-\pi}^{\pi}\tilde{H}(k)dk.
\end{equation}
Here, $\tilde{H}(k)$ refers to the reduced Hamiltonian in the $k$-space.
There is the particle-hole symmetry, in other words,  $\tilde{H}(-k)=\tilde{H}(k)$. So $\tilde{H}(k)$ reads
\begin{eqnarray}
\tilde{H}(k)&=&\frac{B}{2}(c^+_kc_k+c^+_{-k}c_{-k})
\\ \nonumber &-& j\sin k(Cc^+_kc^+_{-k}-C^*c_{-k}c_k)- F \cos k (c^+_kc_k+c^+_{-k}c_{-k})
\\ \nonumber&=&(\frac{B}{2}-F\cos k )(c^+_kc_k+c^+_{-k}c_{-k})-j\sin k  (Cc^+_kc^+_{-k}-C^*c_{-k}c_k).
\end{eqnarray}
We choose four basic vectors $|0>_{k}|0>_{-k},|1>_{k}|1>_{-k},|1>_{k}|0>_{-k}$ and $|0>_{k}|1>_{-k}.$ For simplicity, we let
$ f(k)\equiv \frac{B}{2}-F\cos k$ and $g(k)\equiv C\sin k=|g(k)|e^{j\omega_C}.$  The $\omega_C$ is the phase angle of the complex number $g(k)$\cite{Liao2020}. {So $\tilde{H}(k)$ reads
 \begin{equation}
 \tilde{H}(k)=\left( \begin{matrix}
 0&jg^{*}(k) &0&0
 \\-jg(k)&2f(k)&0&0
 \\0&0&f(k)&0
 \\0&0&0&f(k)
\end{matrix}\right)=f(k)I_4+\left( \begin{matrix}
-f(k)&jg^{*}(k) &0&0
 \\-jg(k)&f(k)&0&0
 \\0&0&0&0
 \\0&0&0&0
\end{matrix}\right).
\label{eq:matrix}
\end{equation}}

The four eigenvalues of the energy are
\begin{equation}
E_m(k)=f(k)\pm\sqrt{f^2(k)+g(k)g^*(k)};f(k);f(k), m=0,1,2,3.
\end{equation}
 The corresponding eigenvectors $|\psi_m(k)>$ satisfy
 \begin{equation}
 J\tilde{H}(k)|\psi_m(k)>=JE_m(k) |\psi_m(k)>.
 \end{equation}
When the exchange couple $J<0$, $ JE_0(k)=J[f(k)+\sqrt{f^2(k)+|g(k)|^2}]$ is ground state energy. When the exchange couple $J>0$, $ JE_0(k)=J[f(k)-\sqrt{f^2(k)+|g(k)|^2}]$ is ground state energy. {One can choose the zero-point energy being $f(k).$ When the exchange couple $J<0$, the ground state energy $E_g$ is $J\sqrt{f^2(k)+|g(k)|^2}$. When the exchange couple $J<0$, the ground state energy $E_g$ is $-J\sqrt{f^2(k)+|g(k)|^2}$. Ignoring of the different $J$, we introduce the reduced energy $\epsilon(k)$ to describe the energy dispersion relation of the two branches of wave functions maybe possessing non-trivial Berry phase. It reads
\begin{equation}
\epsilon(k)\equiv\pm \sqrt{f^2(k)+|g(k)|^2}.
\end{equation}
 If one know the values of parameters $R, C$ and $D$ in special  $\Delta$ and $h$ values, one can the energy dispersion relation of $\epsilon(k)$.}
\section{The order parameters and the matrix-product-state method}
\label{sect:MPS}
\begin{figure}
 \centering
 \includegraphics[angle=0,height=6.0cm,width=6.0cm,bbllx=80pt,bblly=134pt,bburx=540pt,bbury=621pt]{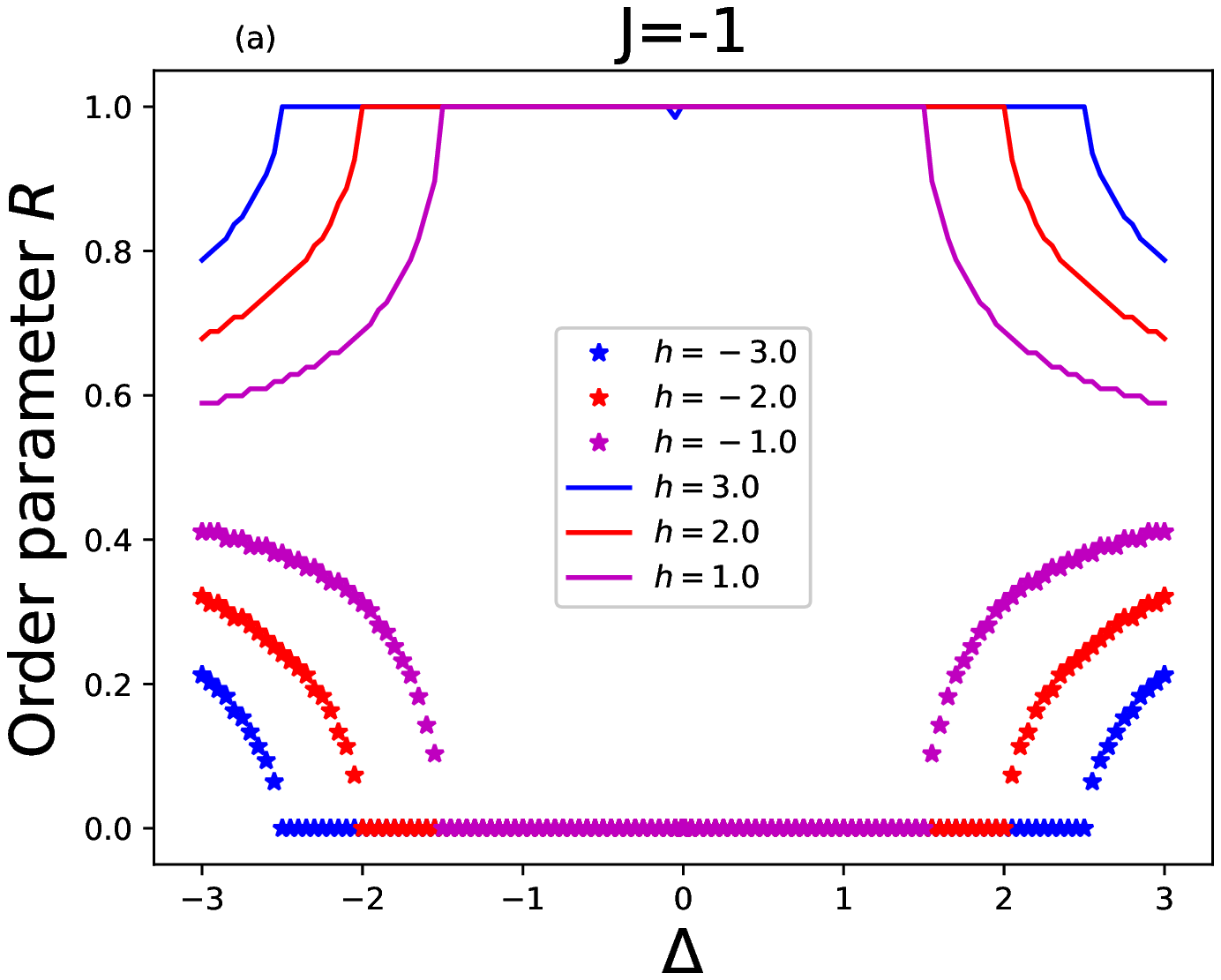}
  \includegraphics[angle=0,height=6.0cm,width=6.0cm,bbllx=80pt,bblly=134pt,bburx=540pt,bbury=621pt]{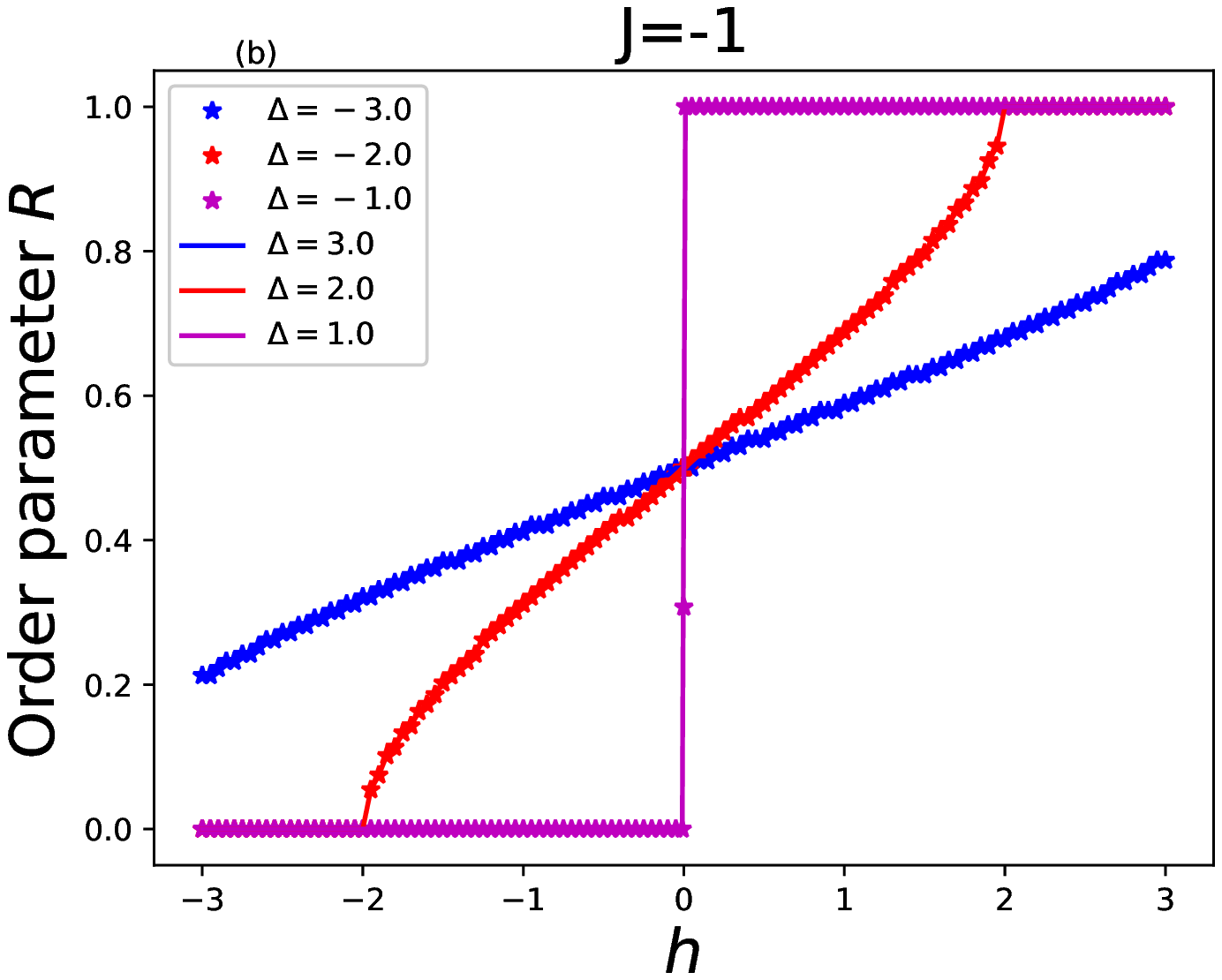}
  \includegraphics[angle=0,height=6.0cm,width=6.0cm,bbllx=80pt,bblly=134pt,bburx=540pt,bbury=621pt]{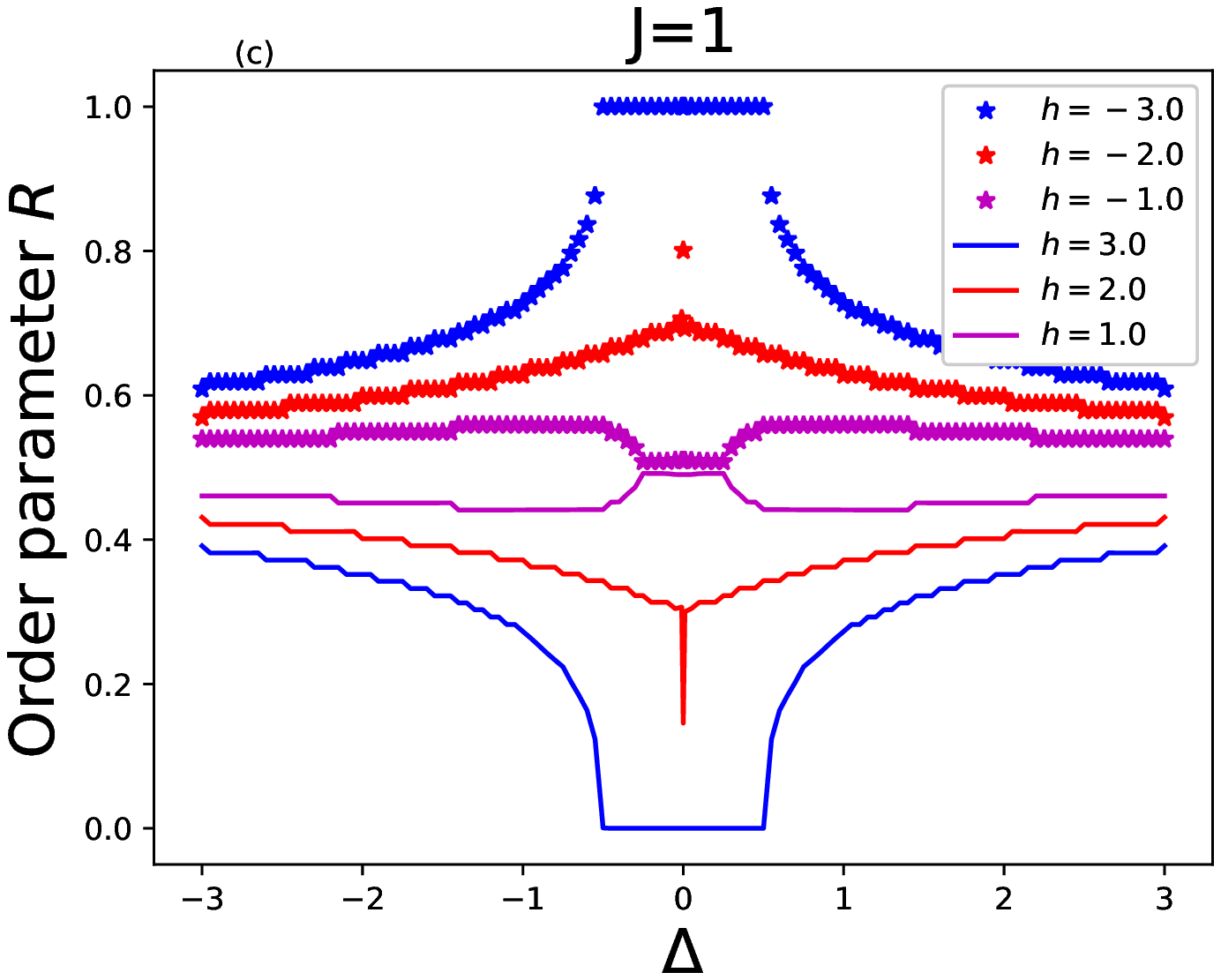}
  \includegraphics[angle=0,height=6.0cm,width=6.0cm,bbllx=80pt,bblly=134pt,bburx=540pt,bbury=621pt]{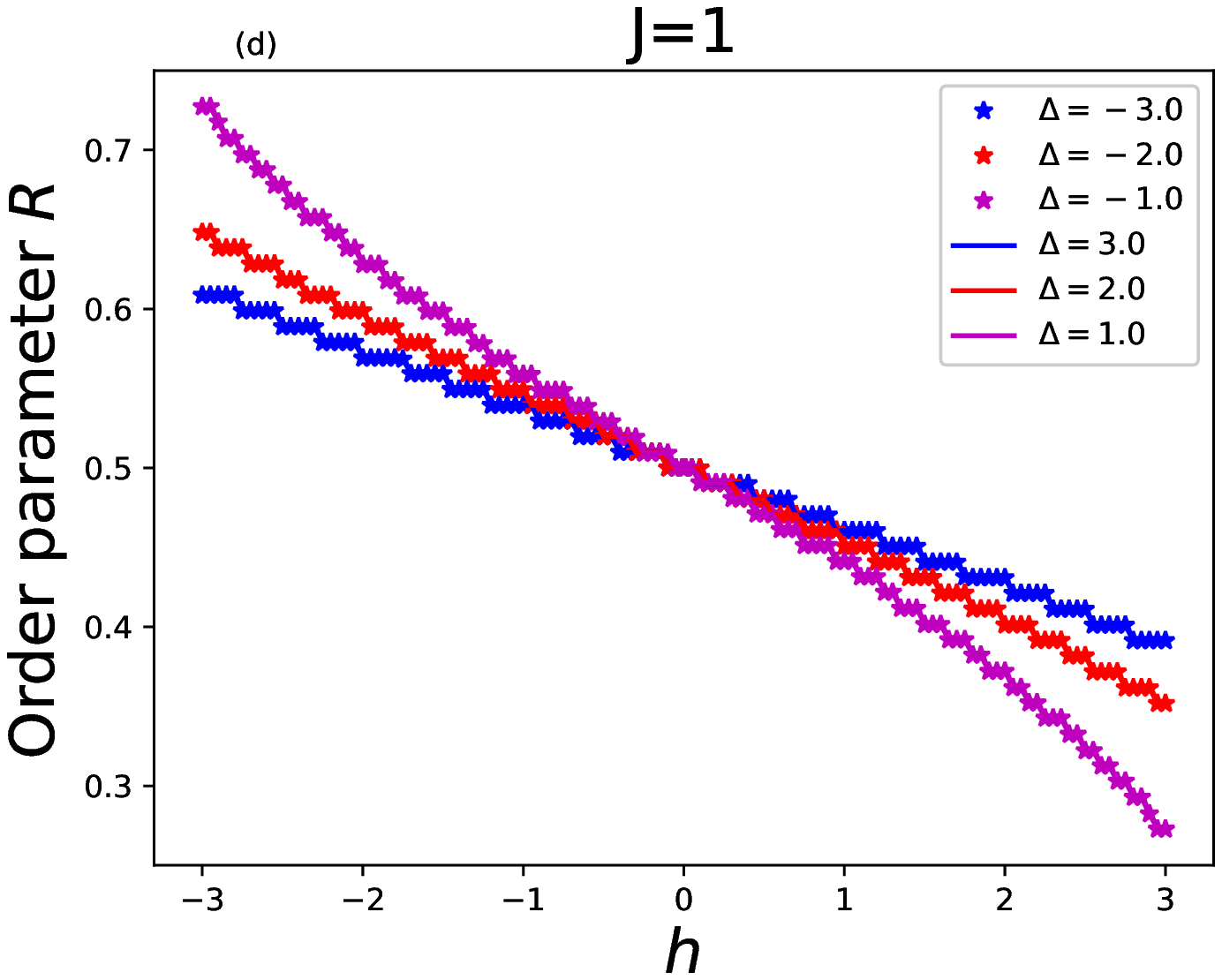}
    \caption{The order parameter $R$ dependence of the reduced field strength $h\in[-3,3]$ and the spin anisotropy parameter $\Delta\in[-3,3]$. Here, the lattice number $N=100$.}
\label{Fig:R}
\end{figure}
\begin{figure}
 \centering
 \includegraphics[angle=0,height=6.0cm,width=6.0cm,bbllx=80pt,bblly=134pt,bburx=540pt,bbury=621pt]{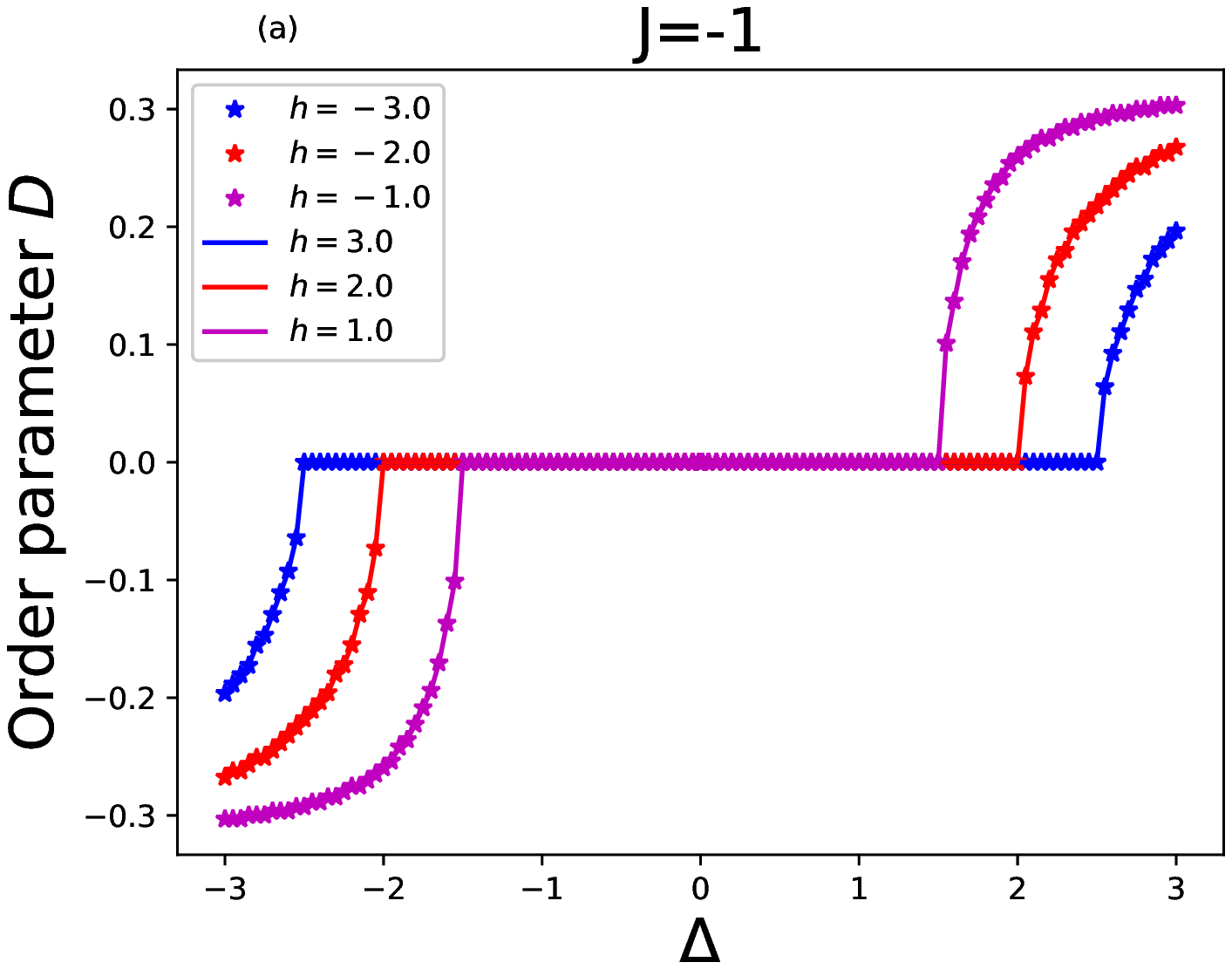}
  \includegraphics[angle=0,height=6.0cm,width=6.0cm,bbllx=80pt,bblly=134pt,bburx=540pt,bbury=621pt]{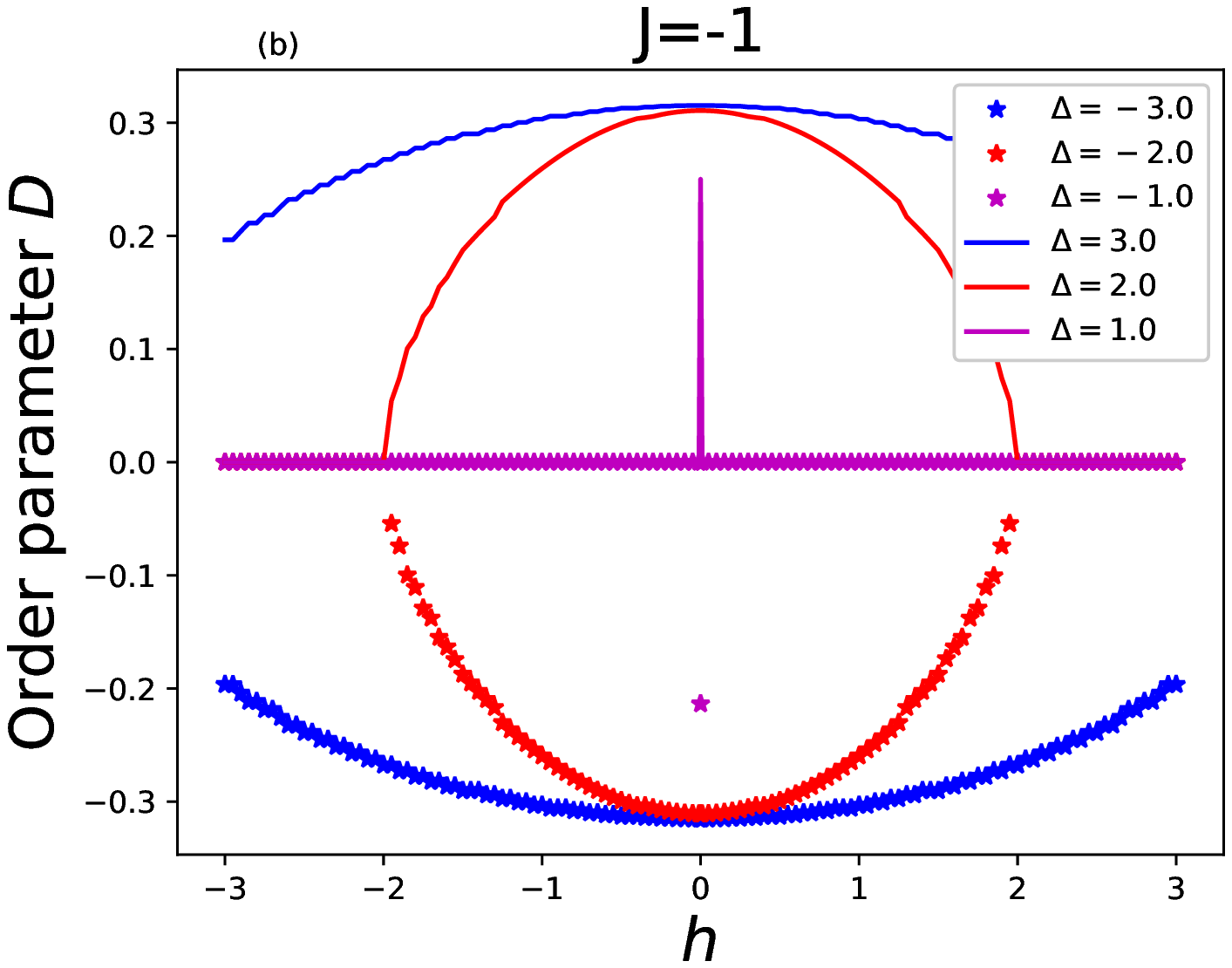}
  \includegraphics[angle=0,height=6.0cm,width=6.0cm,bbllx=80pt,bblly=134pt,bburx=540pt,bbury=621pt]{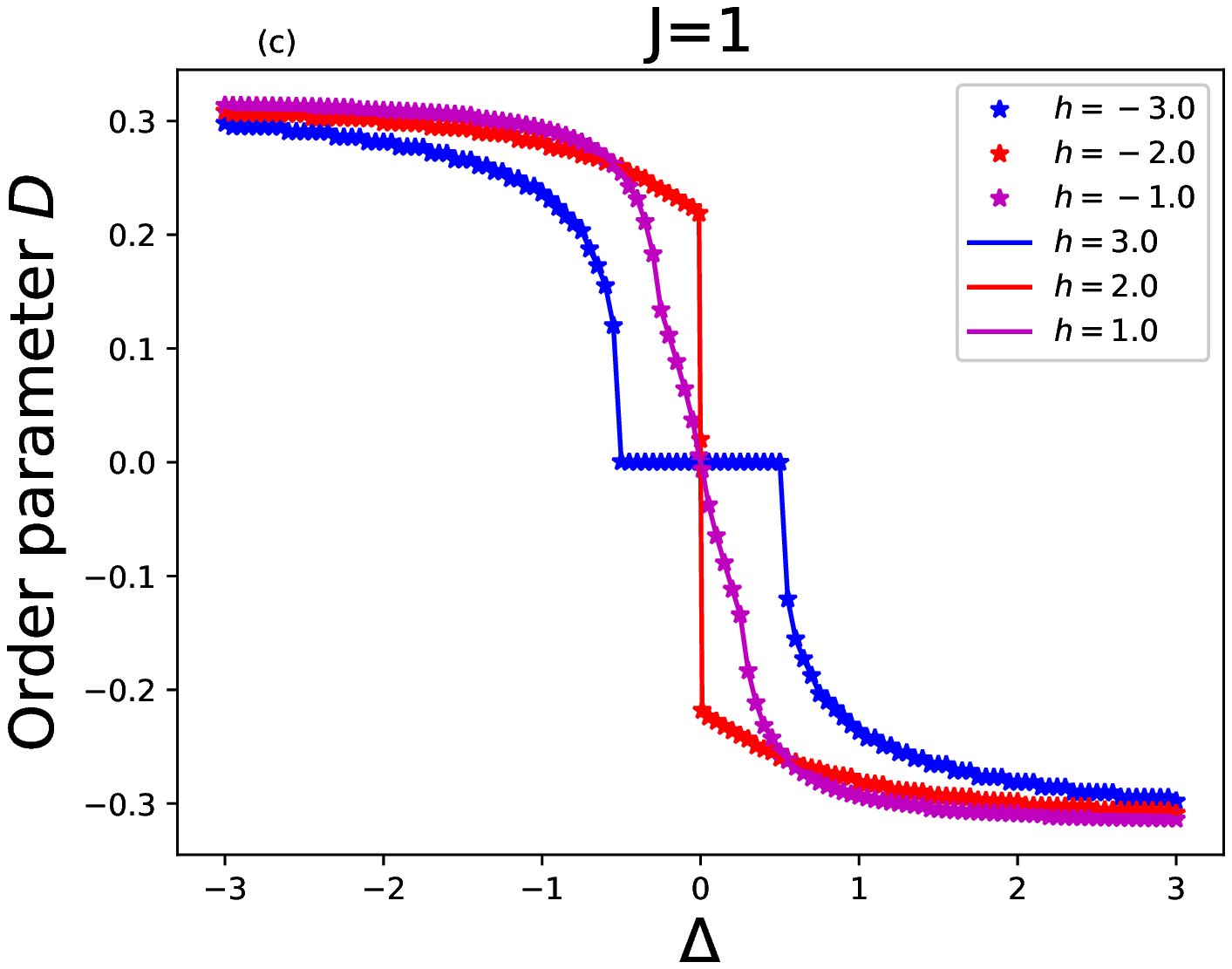}
  \includegraphics[angle=0,height=6.0cm,width=6.0cm,bbllx=80pt,bblly=134pt,bburx=540pt,bbury=621pt]{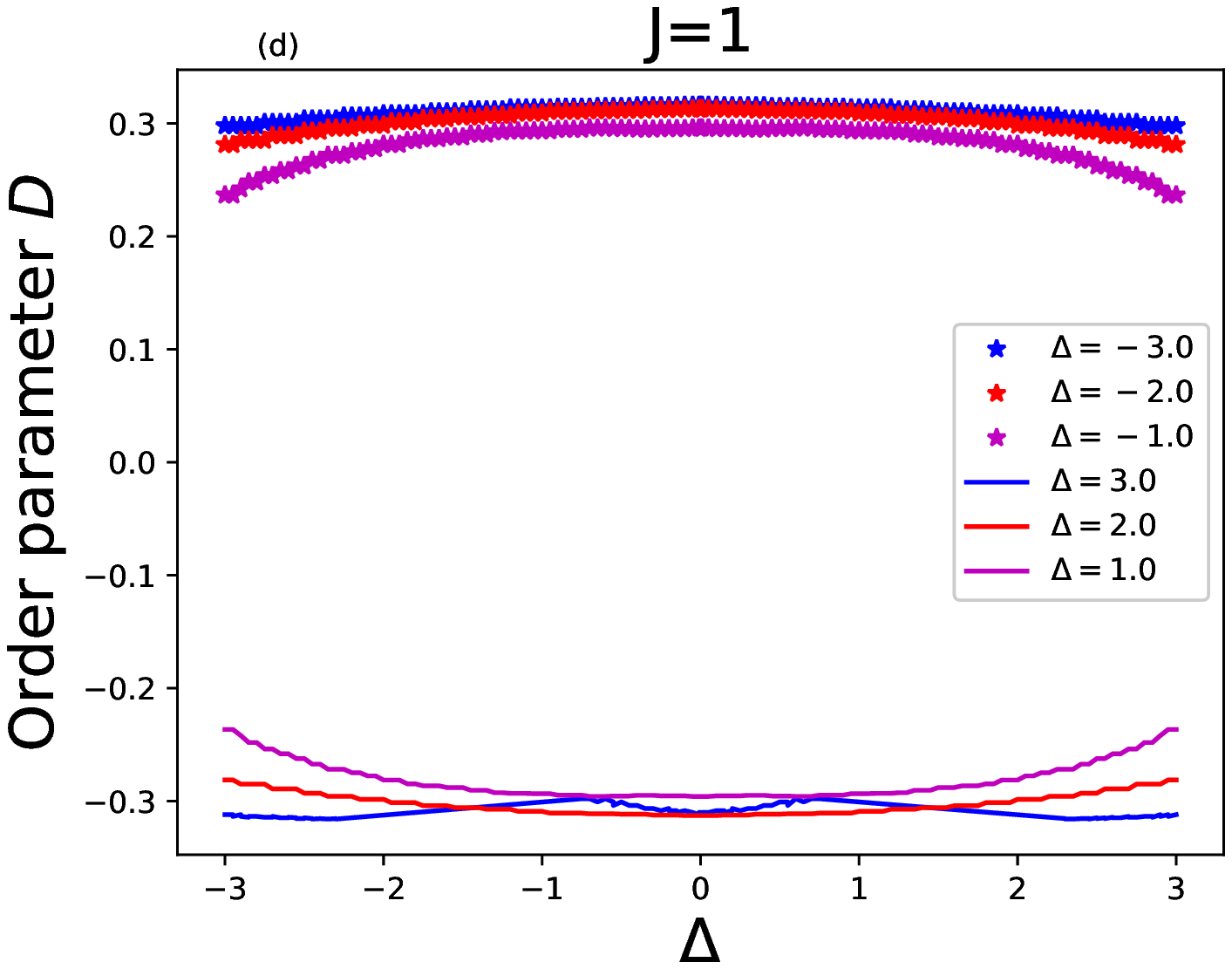}
    \caption{The order parameter $D$ dependence of the reduced field strength $h\in[-3,3]$ and the spin anisotropy parameter $\Delta\in[-3,3]$. Here, the lattice number $N=100$.}
\label{Fig:D}
\end{figure}
\begin{figure}
 \centering
 \includegraphics[angle=0,height=6.0cm,width=6.0cm,bbllx=80pt,bblly=134pt,bburx=540pt,bbury=621pt]{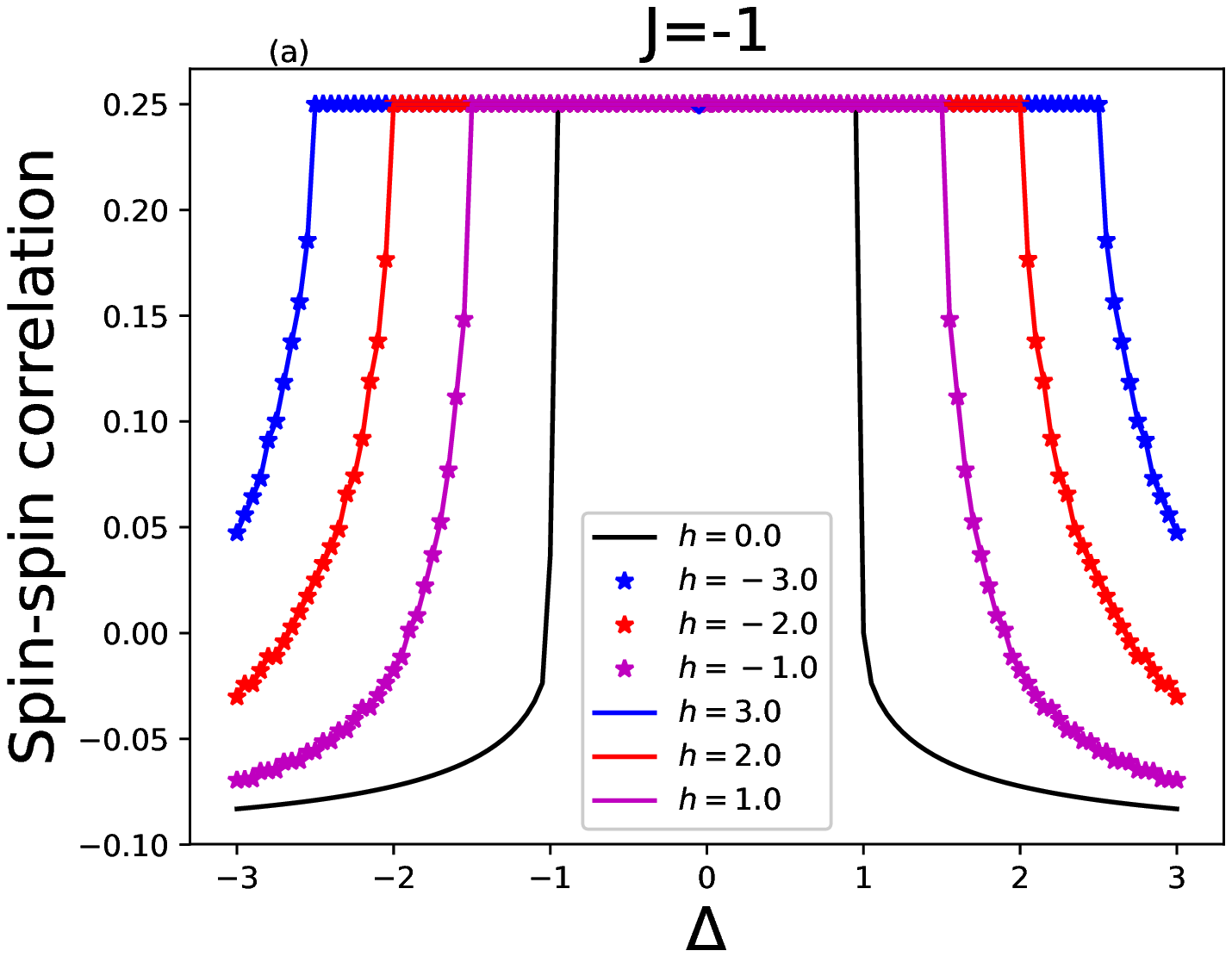}
  \includegraphics[angle=0,height=6.0cm,width=6.0cm,bbllx=80pt,bblly=134pt,bburx=540pt,bbury=621pt]{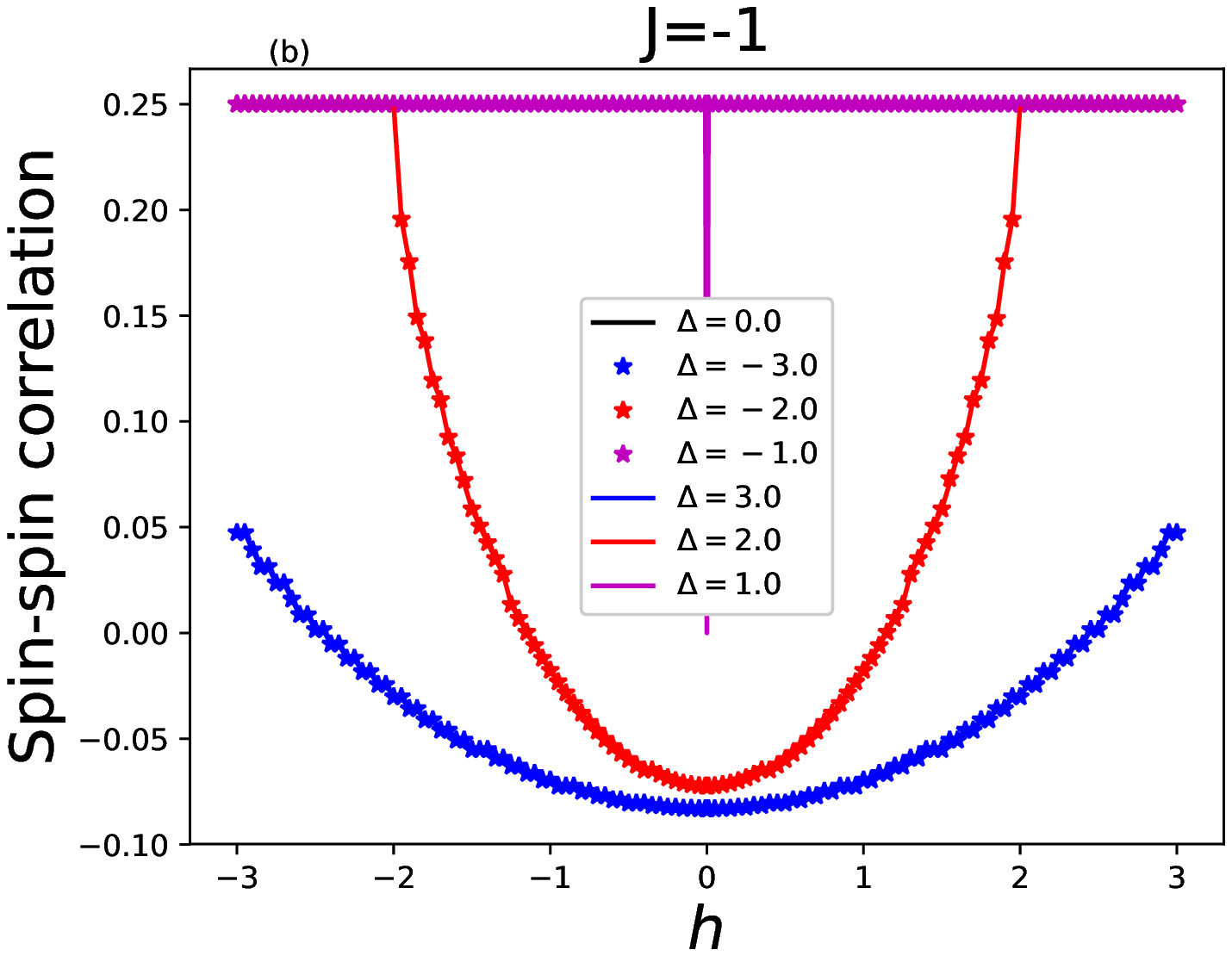}
  \includegraphics[angle=0,height=6.0cm,width=6.0cm,bbllx=80pt,bblly=134pt,bburx=540pt,bbury=621pt]{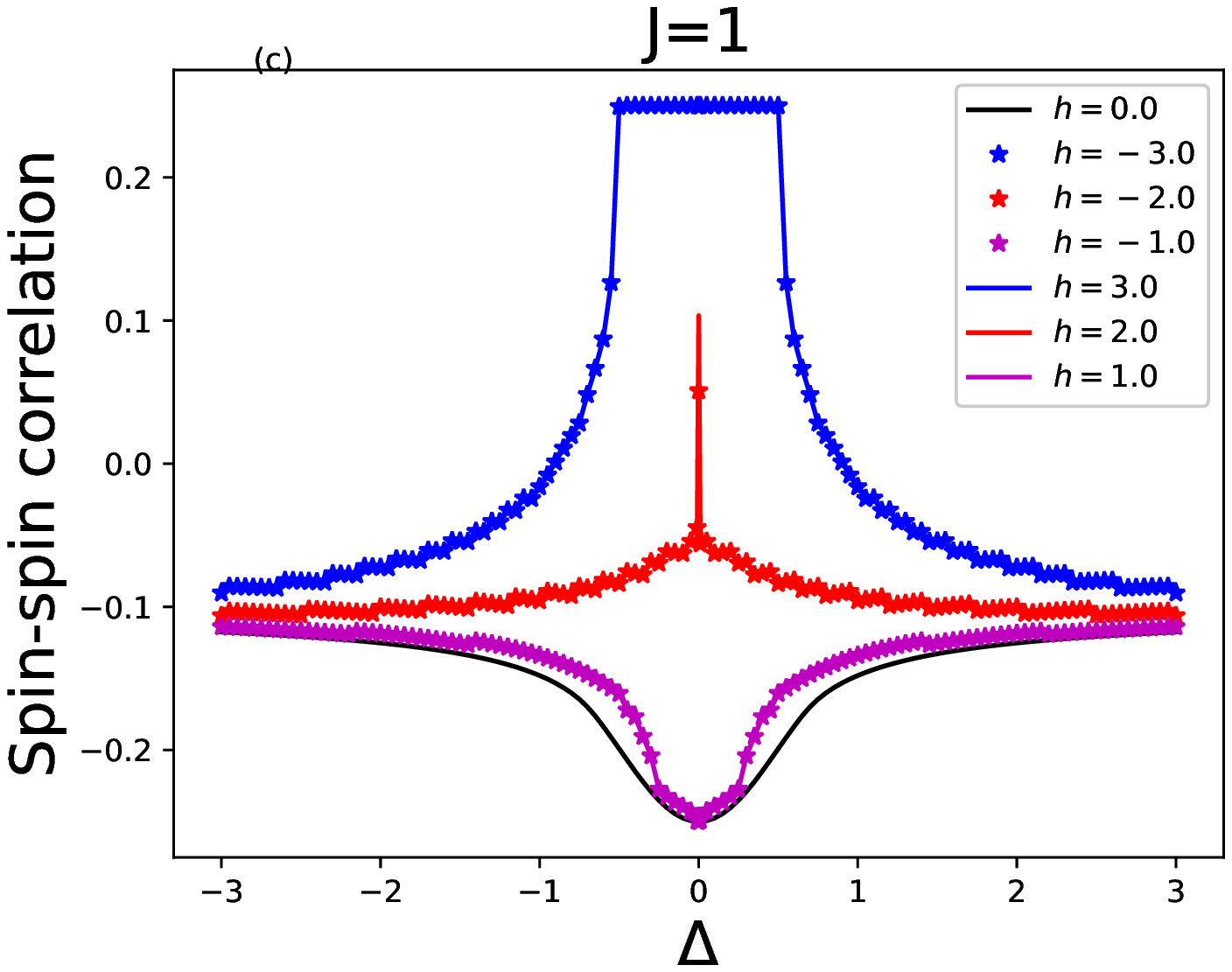}
  \includegraphics[angle=0,height=6.0cm,width=6.0cm,bbllx=80pt,bblly=134pt,bburx=540pt,bbury=621pt]{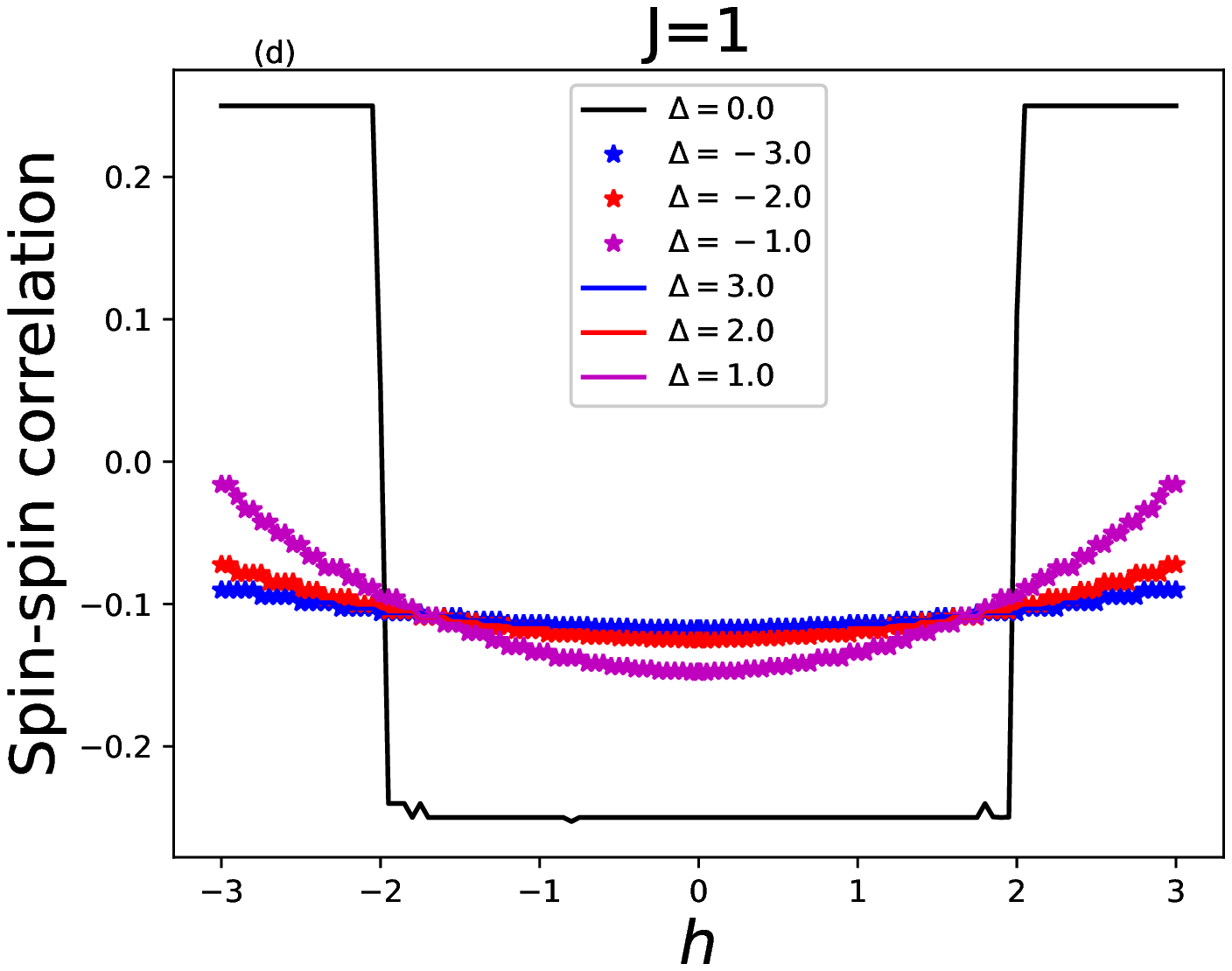}
    \caption{The spin-spin correlation $<S^z_iS^z_{i+1}>$  dependence of the reduced field strength $h\in[-3,3]$ and the spin anisotropy parameter $\Delta\in[-3,3]$. Here, the lattice number $N=100$.}
\label{Fig:spin-spin correlation}
\end{figure}
\begin{figure}
 \centering
 \includegraphics[angle=0,height=6.0cm,width=6.0cm,bbllx=80pt,bblly=134pt,bburx=540pt,bbury=621pt]{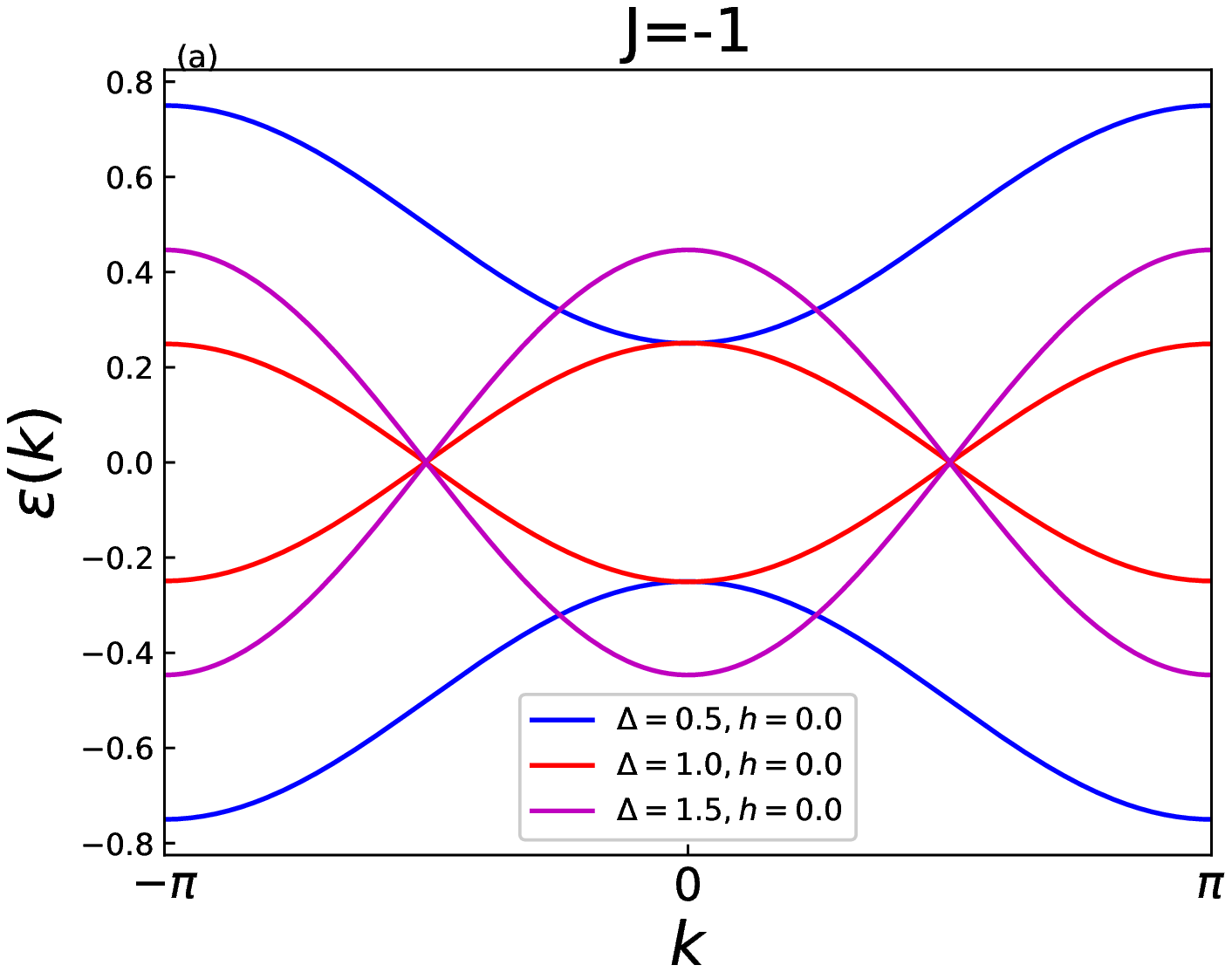}
  \includegraphics[angle=0,height=6.0cm,width=6.0cm,bbllx=80pt,bblly=134pt,bburx=540pt,bbury=621pt]{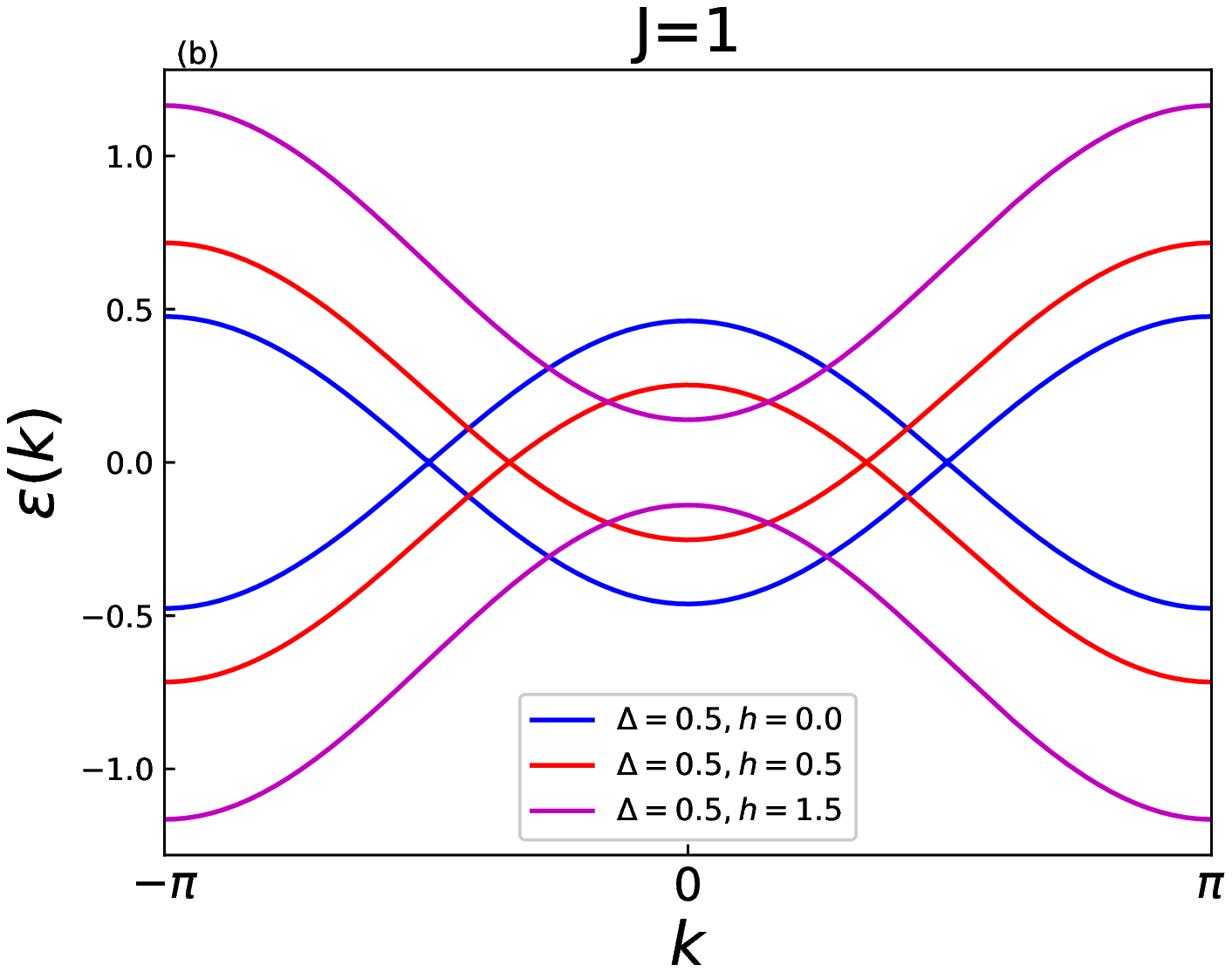}
    \caption{Dispersion relation for the energy $\epsilon(k)$ for different values of $h$ and $\Delta$. The left panel: (a).$J=-1$. $h=0$, $\Delta=0.5, 1, 1.5$. The right panel:(b).$J=1$. $\Delta=0.5$, $h=0, 0.5, 1.5$.}
\label{Fig:dispersion}
\end{figure}
{To thoroughly solve the reduced Hamiltonian, one has to need the values of these order parameters through solving the self-consistent equations . However, to test the validity  of the mean-field approximation, one also has to know the values through another numerical method or the experimental results. In the paper, for} the representative case of $J=-1$ and $J=1$, to calculate the order parameters, we adopt the matrix-product-state algorithm\cite{Vidal,Daley,White,Schollwock}. Matrix product state (MPS) has been a very successful numerical method in studying one-dimensional quantum many-body systems with local and gapped Hamiltonians. We have used a bond dimension up to $\chi=100$ and a singular value decomposition truncation threshold $t=10^{-8}$ for our problem, further checked its convergence by comparing the observables computed with $\chi=150$ for a randomly chosen set of parameters. {MPS algorithm works best for the quantum states with a finite amount of quantum entanglement.} We note that for the special case $\Delta \rightarrow 0$, the model reduces to a classical Ising chain with no quantum entanglement, in which region MPS algorithm is very difficult to converge into the correct ground state. We thus have  discarded this results in this region since they are not reliable. Due to the commutation relation $[\tilde{H}, \overset{N}{\underset{i=1}{\sum}}c^+_ic_{i}]=0$ for the $XXZ$ model, it is noticeable that $C=0$ apart from some special parameter regions where the ground states are degenerate. The magnetization reads $M\equiv <S^z_i>=R-\frac{1}{2}$. There are many literatures about the exact solutions and numerical simulations on the magnetizations and the spin-spin correlations in the $XXZ$ Heisenberg model\cite{Takahashi,Jimbo,Jimbo1996,Kitanine,Kitanine2002,Caux,Caux2005,Giuliano}.  As shown in Fig.\ref{Fig:R} and Fig.\ref{Fig:D}, in the ground state $|g>\rightarrow |\uparrow \cdots \uparrow>$, or $|g>\rightarrow |\downarrow \cdots \downarrow>$,  at the small spin anisotropy parameter $\Delta\in[-1,1]$, the order parameter $R \approx 1$ or $R \approx 0$, and the order parameter $D \approx 0$.  As shown in Fig.\ref{Fig:spin-spin correlation}, The nearest-neighbor spin-spin correlations are drawn which can be used to test the Wick's rule. For the different reduced field strength $h$, there exist the spin anisotropy parameters where order parameters $R$ and $D$ turn discontinuous.  The discontinuities generally reflect the GL quantum phase transition. They also refer to the discontinuous BPs that correspond to the topological quantum transitions. It is noticeable that the order parameter $D$ is not observable. {From Fig. \ref{Fig:R} and Fig. \ref{Fig:D}, one could know the dispersion relation of the energy $\epsilon(k)$. As shown in Fig. \ref{Fig:dispersion}, there are energy crossovers for some parameter $h$ and $\Delta$. There may be a signature of a singular Fermi surface where the gap closes linearly at the critical point. This will induce the nontrivial Berry phase. It means that there exist topological phase transition.}

\section{Berry phase and topological phase diagram}
\label{sect:BP}
For calculating the BP, we can diagonalise this Hamiltonian via a Bogoliubov transformation with two real functions $\theta_k$ and $\phi_k$ satisfying $\theta_{-k}=-\theta_k$ and $\phi_{-k}=\phi_{k}$, yielding
  \begin{equation}
 c_k=\cos\theta_k d_k-je^{j\phi_k}\sin \theta_k d^+_{-k}; c^+_{-k}=-je^{-j\phi_k}\sin \theta_k d_k+\cos \theta_k d^+_{-k}.
 \end{equation}
 {If the non-diagonal terms disappear, one has the condition which reads}
 \begin{equation}
 f(k)\sin (2\theta_k)+[g(k) e^{-j\phi_k}\sin^2\theta_k-g^*(k)e^{j\phi_k}\cos^2\theta_k]=0.
 \end{equation}
In the other words, it reads
 \begin{equation}
 f(k)\sin (2\theta_k)=|g(k)| \cos (2\theta_k); \phi_{k}=\omega_C; \theta_{-k}=-\theta_k.
 \end{equation}
 So $\tilde{H}(k)$ {turns into a bilinear Hamitonian, which reads}
\begin{eqnarray}
\tilde{H}(k)&=&f(k)(c^+_kc_k+c^+_{-k}c_{-k})-jg(k)c^+_kc^+_{-k}+jg^*(k)c_{-k}c_k
\\ \nonumber&=&f(k)+\sqrt{f^2(k)+|g(k)|^2}(d^+_kd_k+d^+_{-k}d_{-k}-1)
\\ \nonumber&=&f(k)+2\sqrt{f^2(k)+|g(k)|^2}(d^+_kd_k-\frac{1}{2}).
\end{eqnarray}
$|\psi_g> $ is the ground state wave function in the $k$-space. $\phi_{k}$ is a constant. $|\psi_g>$ reads
\begin{eqnarray}
|\psi_g>\equiv \underset{\bigotimes k }{\Pi} |\psi_0(k)>&=&\underset{\bigotimes k }{\Pi}[\cos \theta_k|0>_{k}|0>_{-k}-je^{j\phi_k}\sin \theta_k|1>_{k}|1>_{-k}].
\end{eqnarray}
{Based on the ground state $|\psi_g> $,  one could get the two-point correlators which read
\begin{eqnarray}
<c^+_{k'}c_{k''}>&=&\{n(k') \cos^2 \theta_{k'}+[1-n(k')]\sin^2 \theta_{k'}\}(\delta_{k',k''})
\\ \nonumber &=&\{\frac{1}{2}+[n(k')-\frac{1}{2}]\cos (2\theta_{k'})\}(\delta_{k',k''});
\\ \nonumber <c_{k'}c_{k''}>&=&\{[n(k')-\frac{1}{2}]je^{j\phi_{k'}}\sin (2\theta_{k'}) \}(\delta_{k',-k''}).
\label{eq:cc}
\end{eqnarray}
Here, the Fermi distribution function reads
\begin{equation}
n(k)=\frac{1}{1+\exp(-2\beta \sqrt{f^2(k)+|g(k)|^2} )}, \beta=\frac{1}{k_BT}.
\end{equation}
When $T\rightarrow 0$, we think $n(k')=1$. }

\begin{figure}
 \centering
 \includegraphics[angle=0,height=6.0cm,width=6.0cm,bbllx=80pt,bblly=134pt,bburx=540pt,bbury=621pt]{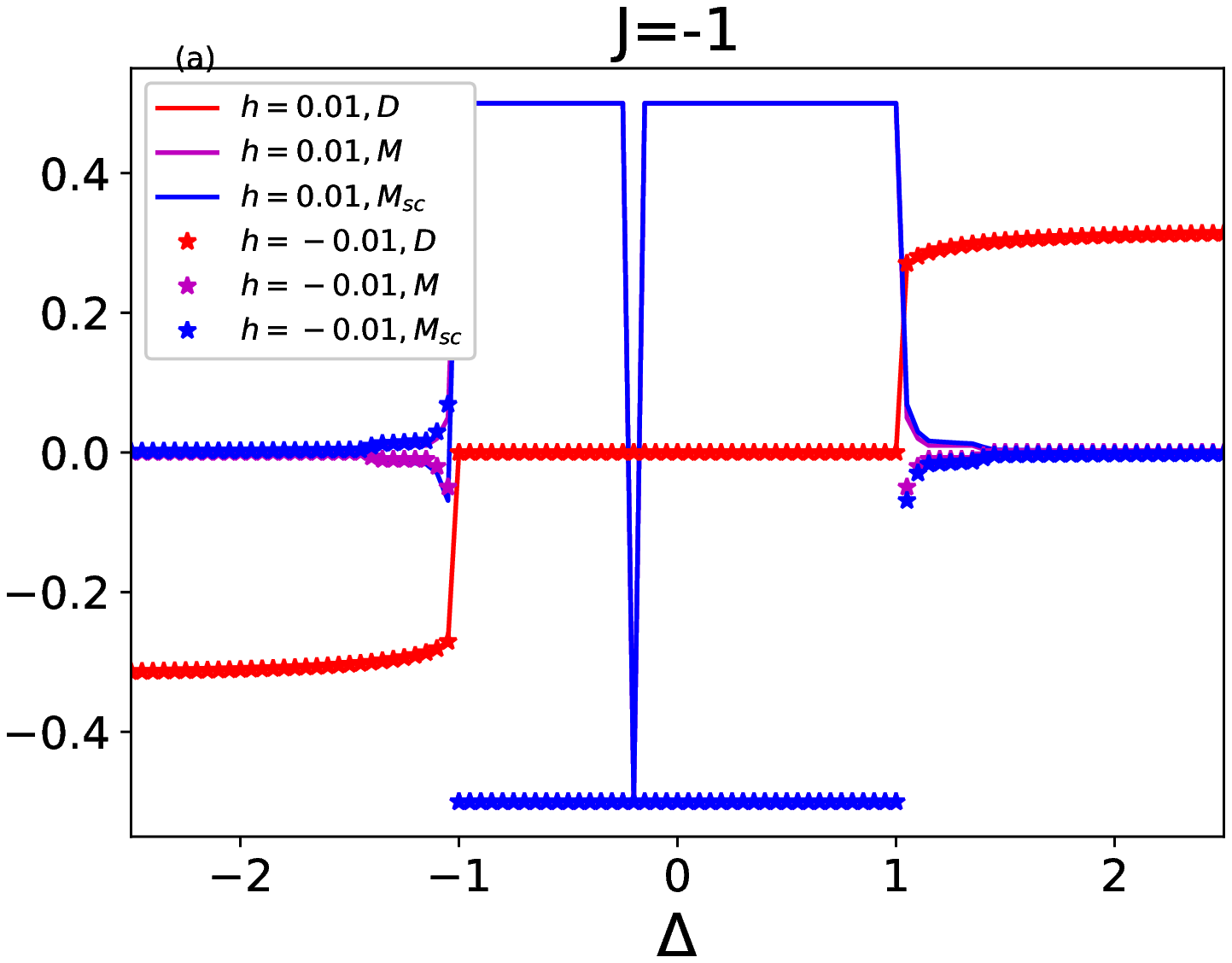}
  \includegraphics[angle=0,height=6.0cm,width=6.0cm,bbllx=80pt,bblly=134pt,bburx=540pt,bbury=621pt]{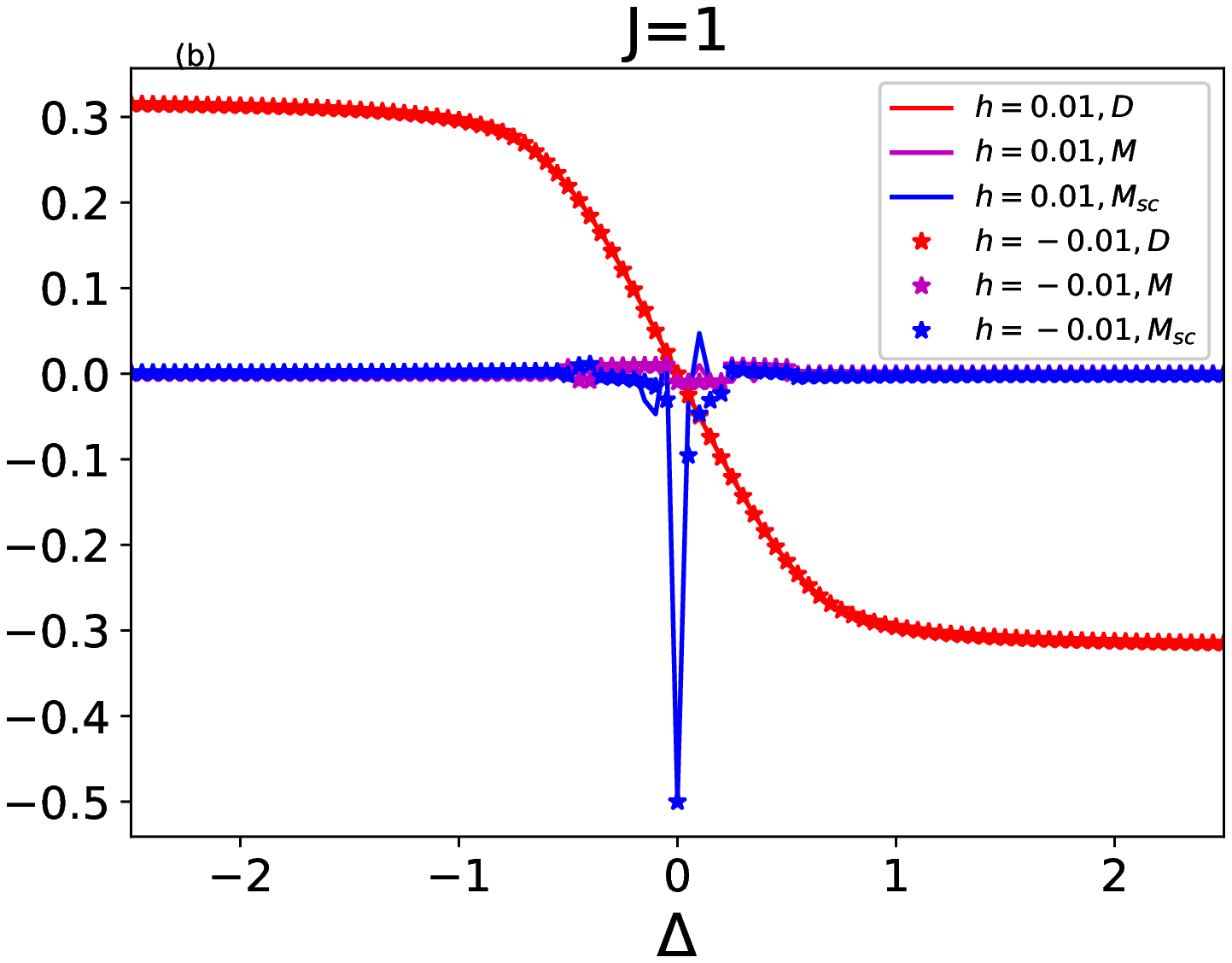}
    \caption{The direct results $M$ compared with the ones $M_{_{SC}}$ in the mean-field theory when $h=\pm 0.01$ and $\Delta\in[-2.5.2.5]$. The left panel: (a).$J=-1$. The right panel:(b).$J=1$.}
\label{Fig:compared}
\end{figure}

{Based on the Eq.(\ref{eq:cc})}, the order parameters satisfy the self-consistent equation which reads\cite{Liao2020}
\begin{eqnarray}
f(k)&=&R+\frac{h-1}{2}-(D-\frac{\Delta}{2}) \cos k;
\\ \nonumber |g(k)|&=&|C| \sin k;
\\ \nonumber R&=& \frac{1}{2}+\frac{1}{\pi }\int_{0}^{\pi}\frac{[n(k)-\frac{1}{2}]f(k)dk}{\sqrt{f^2(k)+|g(k)|^2}};
\\ \nonumber |C|&=& \frac{1}{\pi }\int_{0}^{\pi} \frac{[n(k)-\frac{1}{2}]|g(k)|\sin kdk}{\sqrt{f^2(k)+|g(k)|^2}};
\\ \nonumber D&=& \frac{1}{\pi }\int_{0}^{\pi} \frac{[n(k)-\frac{1}{2}]f(k)\cos kdk}{\sqrt{f^2(k)+|g(k)|^2}}.
\label{eq:self-consitent}
\end{eqnarray}
{There are some couples between three self-consistent parameters and some piecewise function in Eq.(\ref{eq:self-consitent}) . Those make the iterations very complex. And it is hard to obtain the convergent solutions. Since one already know the parameters through MPS method, one could introduce some artifices to test the mean-field approximation. In other words,
one can rewrite} the self-consistent solution for the magnetization $M_{_{SC}}$ which reads
\begin{eqnarray}
M_{_{SC}} &\equiv &R_{_{SC}}-\frac{1}{2}=\frac{1}{2\pi }\int_{0}^{\pi}\frac{f(k)dk}{\sqrt{f^2(k)+|g(k)|^2}}
\\ \nonumber &=&\frac{1}{2\pi }\int_{0}^{\pi} \frac{[2M+h+(\Delta-2D)\cos k]dk}{\sqrt{[2M+h+(\Delta-2D)\cos k]^2+4|C|^2\sin^2 k}}.
\end{eqnarray}
{In the right part of above equation, one can use the parameter values solved through MPS method.}
When $|C|=0^+$,
\begin{equation}
\operatorname{if} |\Delta-2D|\geq |2M+h|, M_{_{SC}}=-\frac{1}{2}+\frac{1}{\pi}\arccos(\frac{2M+h}{2D-\Delta}).
\end{equation}
If $|\Delta-2D|\leq -(2M+h), M_{_{SC}}=-\frac{1}{2}$. And if $|\Delta-2D|\leq (2M+h), M_{_{SC}}=\frac{1}{2}.$
The validity of the mean-field theory could be tested through the relationship between the self-consistent solutions and the numerical results. Making the comparison $M$ with $M_{_{SC}}$ at different parameters $\Delta$ and $h$, we could know the region where the approximation is established. As an example when $h\approx 0$ meaning that the LF vanishes, the compared results show that the approximation is applicable in the Fig.\ref{Fig:compared}.

The BP of the ground state is defined  by
 \begin{equation}
 \gamma_g=j\int_{-\pi}^{\pi}<\psi_0(k)|\frac{d}{dk}|\psi_0(k)>dk,f(k)\sin (2\theta_k)=|g(k)| \cos (2\theta_k).
 \end{equation}
 We map the Hamiltonian to a two-level system and have a close curve $\partial \Omega$ where the point $(x ,y)$ satisfies\cite{Liao,Liao2020}
\begin{equation}
(\frac{x-B/2}{F})^2+(\frac{y}{|C|})^2=1, \frac{B}{2}=M+\frac{h}{2}; F=D-\frac{\Delta}{2}.
\end{equation}
 Because $\omega_{C}$ is a constant which is independent of $k$, the criterion for nonzero BP is decided by the relation between the point $(0,0)$ and the curve $\partial \Omega$ \cite{Liao}. In other words, it depends on the size of the relationship between $|B|/2$ and $|F|$. The BP reads
\begin{equation}
\gamma_g=\frac{\operatorname{sgn}(J)[\operatorname{sgn}(|F|-|B|/2)+1]\pi}{2}=\frac{\operatorname{sgn}(J)[\operatorname{sgn}(|D-\frac{\Delta}{2}|-|M+\frac{h}{2}|)+1]\pi}{2}.
\label{eq:g}
\end{equation}
Here $\operatorname{sgn}(\zeta>0)=1; \operatorname{sgn}(\zeta=0)=0; \operatorname{sgn}(\zeta<0)=-1.$

\begin{figure}
 \centering
 \includegraphics[angle=0,height=6.0cm,width=6.0cm,bbllx=80pt,bblly=134pt,bburx=540pt,bbury=621pt]{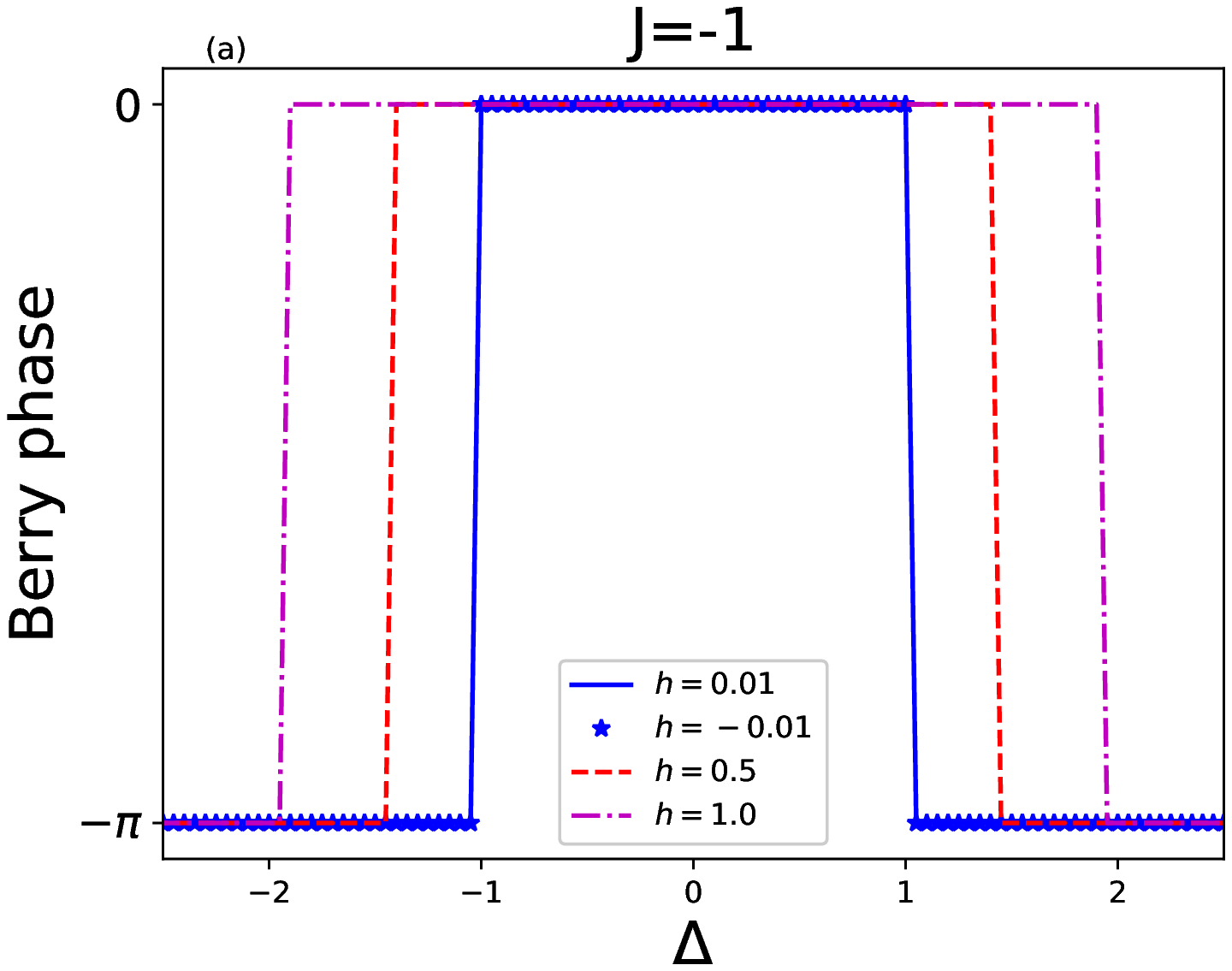}
  \includegraphics[angle=0,height=6.0cm,width=6.0cm,bbllx=80pt,bblly=134pt,bburx=540pt,bbury=621pt]{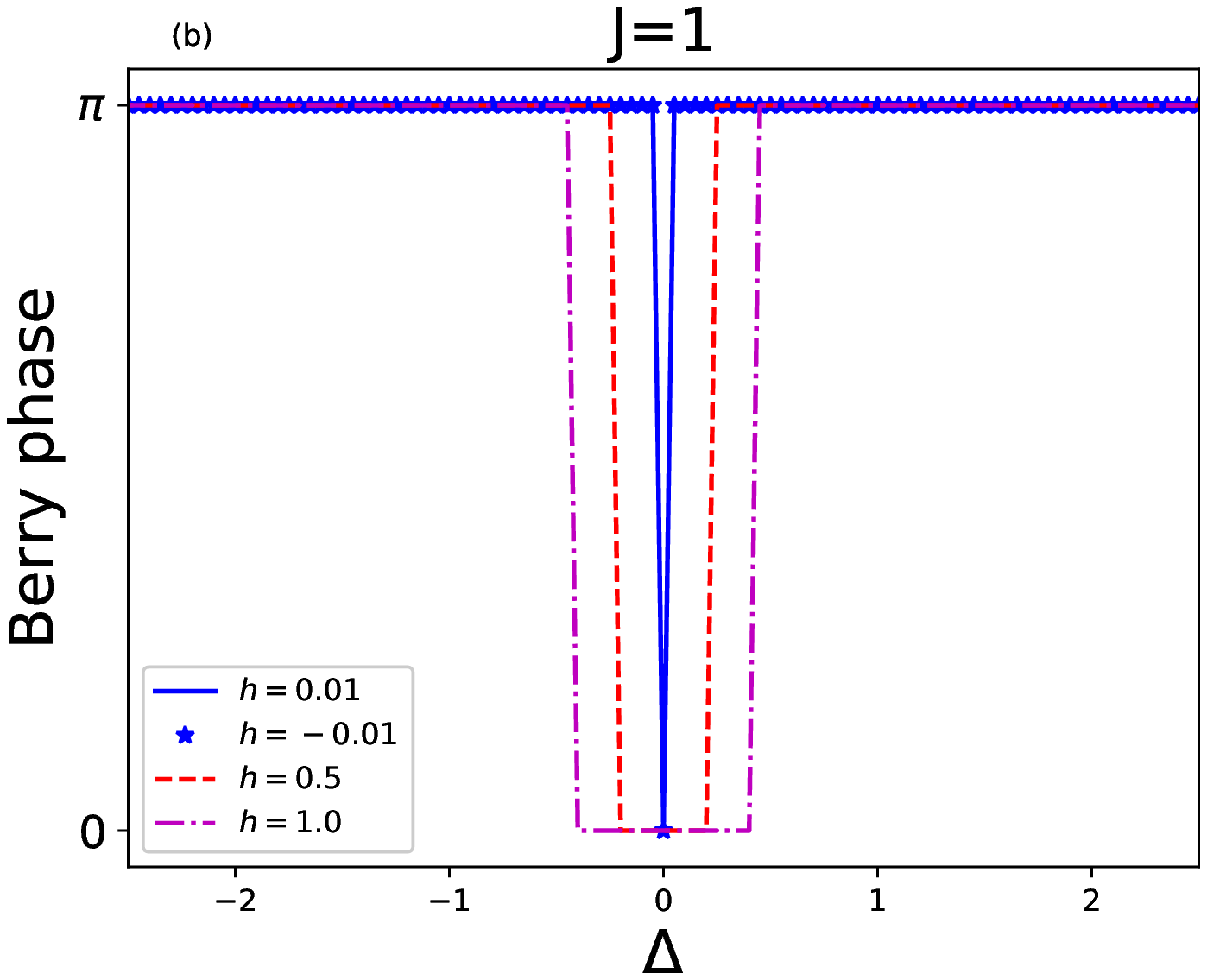}
    \caption{The Berry phase $\gamma_g$  dependence of the spin anisotropy parameter $\Delta \in[-2.5.2.5]$ in the different values of $h=\pm 0.01$, $h=0.5$ and $h=1.0$. Discard the results in this region $\Delta \rightarrow 0$ where MPS algorithm is not reliable. The left panel: (a).$J=-1$. The right panel:(b).$J=1$. }
\label{Fig:BP}
\end{figure}

\begin{figure}
 \centering
 \includegraphics[angle=0,height=6.0cm,width=6.0cm,bbllx=80pt,bblly=134pt,bburx=540pt,bbury=621pt]{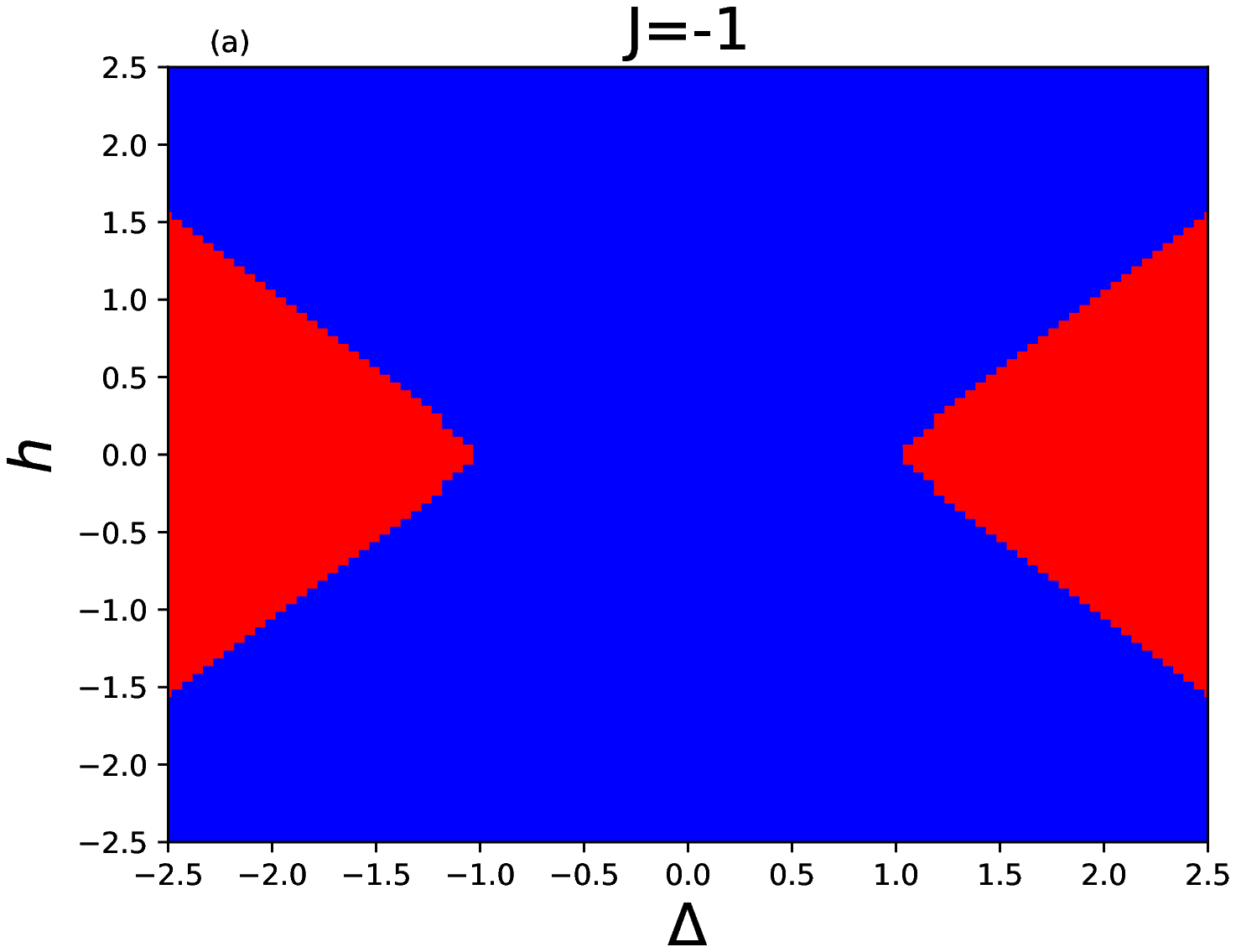}
 \includegraphics[angle=0,height=6.0cm,width=6.0cm,bbllx=80pt,bblly=134pt,bburx=540pt,bbury=621pt]{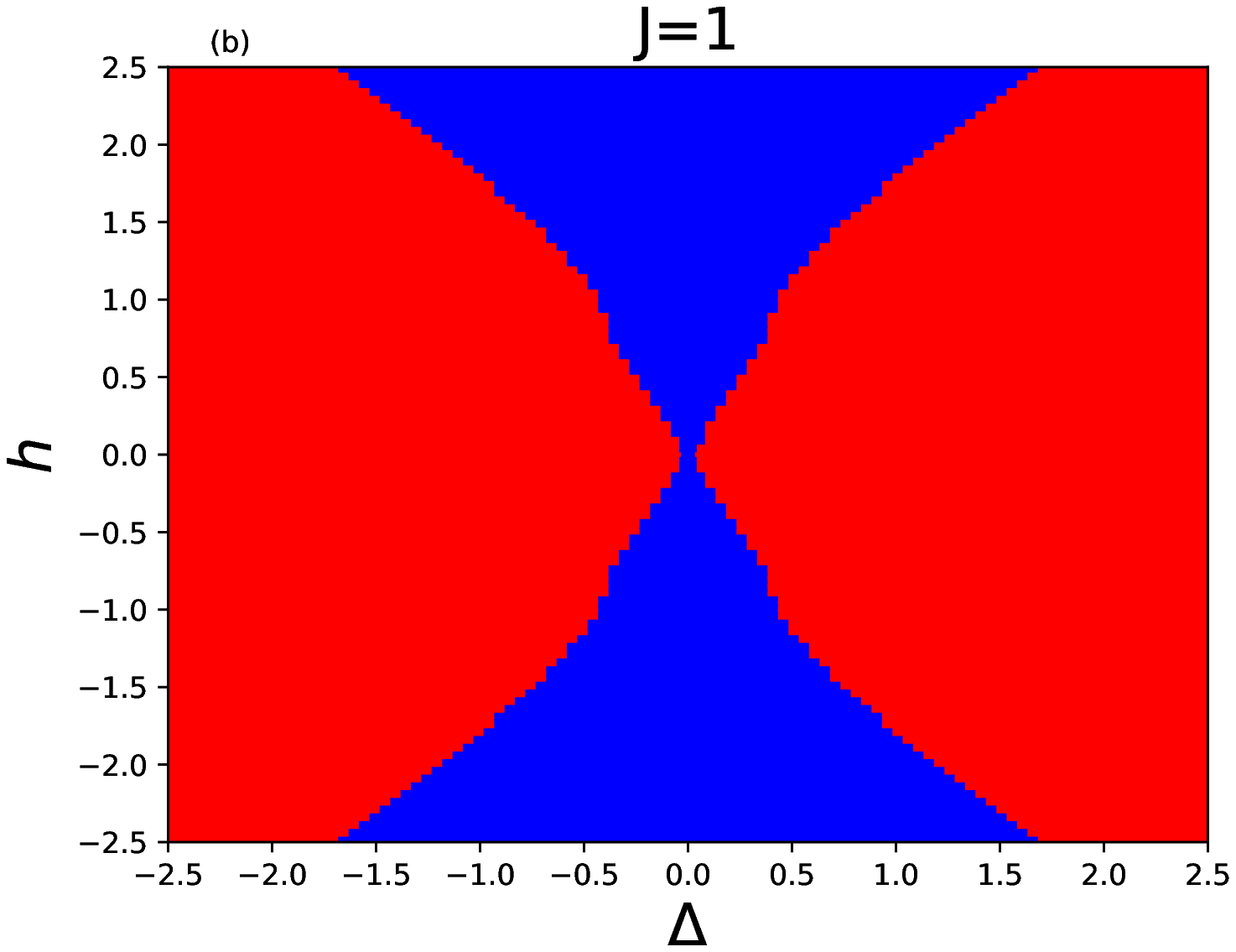}
    \caption{The topological phase diagram dependence of the reduced field strength $h\in[-2.5.2.5]$ and the spin anisotropy parameter $\Delta\in[-2.5,2.5]$. Discard the results in this region $\Delta \rightarrow 0$ where MPS is not reliable. The left panel: (a).$J=-1$. The red, black and blue point indicates $\gamma_g=-\pi,-\frac{\pi}{2}, 0$.
    The right panel: (b).$J=1$. The red, black and blue point indicates $\gamma_g=\pi,\frac{\pi}{2}, 0$.}
\label{Fig:digram}
\end{figure}

The BPs dependence of the spin anisotropy parameter $\Delta$ and the reduced field strength $h$ are shown as the Fig.\ref{Fig:BP} and Fig.\ref{Fig:digram}. Due to the ground-state degeneracy in some special points, it should be noticed that the BP is not well defined if the ground states have different BPs. And on the phase boundary, there exist some inaccurate BP predictions even though the mean-field approximation works. It need be pointed that the phase diagrams are not right when $\Delta\rightarrow 0$ where the MPS method is not applicable {in the classical Ising model where there are not quantum entanglements.} For example, when $\Delta\rightarrow 0$ and $h\rightarrow 0$ , the $XXZ$ turns into Ising model, there exists the nonzero BP $\gamma_g=\pi$ in the case $J>0$ \cite{Liao2020}. This conclusion shown in the Fig.\ref{Fig:BP}, that claims the BP $\gamma_g=0$ when $\Delta=0$ and $h=\pm0.01$ in the case $J>0$, is not right.
\section{Results and discussion}
\label{sect:Results}
In conclusion, we have analyzed the spin-$1/2$ $XXZ$ model in the one-dimensional lattice using the mean-field approximation based on the Wick's theorem. The validity of the approximation is tested through comparing the result calculated by the MPS algorithm. The topological phase diagrams characterized by the BP show the different parameter influences on the symmetries of ground states. As shown in Fig.\ref{Fig:digram}, for the case $J<0$, the phase boundary is nearly linear, and for the case $J>0$,  the phase boundary is nearly hyperbolic. When the LF vanishes in Fig.\ref{Fig:BP}, for the case $J<0$, at the small spin anisotropy parameter, the BP of the ground state keeps nought. For the case $J>0$, there exist nonzero BPs at all spin anisotropy parameters.

The present calculation for the BP of the one-dimensional $XXZ$ model enlarges  understanding the topological quantum transition in the spin system. Compared with the phase diagram in the Ref.\cite{Cheng} characterized by the winding number as the topological order parameter in the $XY$ model, the phase diagram is more complicated but keeps conformal. {Our work practically implies the jumping of a quantity being not observable plays an important role in determining the topological transition.}

The topological quantity is a new order parameter explaining the phase transition. In general, the  topological quantum phase diagrams are mutually independent of the ordinary GL quantum phase diagrams. Even through both share the same transition points in some special systems, such as the one-dimensional transverse field Ising model, there are not necessary relevances between the topological order parameter, such as the BP, and the ordinary order parameter, such as the quantum magnetic susceptibility.

The Berry phase, as a well-known topological order parameter, it can be well used to determine the phase transition. But it is only a mathematical tool. In the topological phase, what is the physical property such as quantum entanglement and topological entanglement entropy? It still keeps an open question.

\section*{Acknowledgments}
This work was supported by the National Basic Research Program of China under Grant No. 2016YFA0301903; the National Natural Science Foundation of China under Grants No. 61632021 No. 11574398, No. 11174370, No. 11574398, and No. Y6GJ161001.

\end{document}